\documentclass[
  ,draft            
  ]
  {aipproc}

\layoutstyle{6x9}

\begin{document}

\title{On the Growth of Dust Particles in a
Non-Uniform Solar Nebula}

\author{Nader Haghighipour}{
  address={Department of Terrestrial Magnetism and 
NASA Astrobiology Institute, \\Carnegie Institution of
Washington, 5241 Broad Branch Road, Washington, DC 20015}
}

\begin{abstract}
A summary of the results of a numerical study of the growth
of solid particles in the vicinity of an azimuthally
symmetric density enhancement of a protostellar disk 
are presented. The effects of gas drag and pressure
gradients on the rate of growth of dust particles
and their settling on the midplane of the nebula
are also discussed.
\end{abstract}

\maketitle


\section{INTRODUCTION}

It has recently been shown that 
the combined effect of gas drag and pressure gradients
causes solid objects to rapidly migrate toward the location
of maximum pressure in a gaseous nebula \cite{hb03a,hb03b}.
Such rapid migrations result in the accumulation
of objects in the vicinity of local pressure enhancement
and may increase the rates of their collisions and
coagulations.
In a disk with a mixture of gas and
small solid particles, the motion of an object
is also affected by its interaction with the
particulate background.
An object may, in such an environment, sweep up smaller particles and grow
larger during its gas-drag induced migration. 
In this paper I present a summary of a study of the 
growth-rates of dust grains to centimeter-sized objects in a 
nebula with a local pressure enhancement.

\section{The Model Nebula}

An isothermal and turbulence-free protostellar disk is considered here 
with a solar-type star at its center. The nebula is assumed 
to be a mixture of pure molecular hydrogen at hydrostatic
equilibrium,  and small submicron-sized solid particles.
The density of the gas in this nebula is considered to be
\begin{equation}
{\rho_g}(r,z)\,= \,{\rho_0}\,
\>{\rm {exp}}
\Biggl\{{{GM{m_H}}\over {{K_B}T}}\,
\biggl[{1\over {{({r^2}+{z^2})^{1/2}}}}-{1\over r}\biggr]
-\beta\Bigl({r \over{r_m}}-1 {\Bigr)^2}\Biggr\}.
\end{equation}
\noindent
In equation (1), $M$ is the mass of the central star,
$G$ is the gravitational constant, and ${K_B},T$ and
$m_H$ represent the Boltzmann constant, 
the gas temperature, and the molecular mass of hydrogen, respectively.
The quantities ${\rho_0}\,,{r_m}$ and $\beta$ in equation (1)
are parameters of the system with constant values.
The gas density function
as given by equation (1) ensures that along the vertical
axis, the gravitational attraction of the central star will be
balanced by the vertical component of the pressure gradients
\cite{hb03b}. It also has azimuthally symmetric gas-density-enhanced
regions on any plane parallel to the midplane. On the midplane,
the density of the gas maximizes at $r={r_m}$.

The particles of the background material of a nebula are
strongly coupled to the gas and their motions
are only affected by the gas drag. 
For the submicron-sized particles considered here,
the rate of the gas drag induced migration is so
small that one can assume,
at any position in the nebula, the gas and
particle distribution functions
are proportional. That is, 
${\rho_{\rm {dust}}}(r,z)=f\,{\rho_g}(r,z)$, where 
${\rho_{\rm {dust}}}(r,z)$ is the distribution
function of the background material. The
solid-gas ratio $f$ is taken to be constant and equal to 0.0034.

\section{Equation of Motion}

For an object 
larger than the submicron particles of the background
material, the above-mentioned gas-particle coupling is
less strong. As a result, the object
moves faster than the background particles and may
collide with them. Such collisions may result in
adhesion of the background particles to the moving object
and increase its mass.

The rate of change of the momentum of an object
due to the sweeping up of the background material is proportional
to the rate of the collision of the moving object with those particles. 
Assuming the sticking coefficient is equal to unity,
one can write,
\begin{equation}
{{d{\bf P}}\over {dt}}=\pi\, {\rho_{\rm{dust}}}(r,z)\,
{a^2}\,{{d\ell}\over {dt}}\, ({{\bf V}-{\bf U}}),
\end{equation}
\noindent
where $a$ is the radius of the object and
${d\ell}=|{\bf V}-{\bf U}|\,dt$.
Quantities ${\bf V}$ and ${\bf U}$ in equation (2) represent the
velocities of the object and the particles of the medium.
In writing equation (2), it has been assumed that, 
while sweeping up smaller particles, the
density of the object remains unchanged and it stays perfectly
spherical. It is important to emphasize
that in this equation, the mass of the object, $m$, 
and its radius, $a$, are functions of time,
and are related as
\begin{equation}
{{dm}\over {dt}}=\pi\, {\rho_{\rm{dust}}}(r,z)\,
{a^2}\,{{d\ell}\over {dt}}.
\end{equation}
\noindent
Equation (3) immediately implies
\begin{equation}
{{da}\over {dt}}={1\over 4}
\Bigl({{\rho_{\rm {dust}}}\over \rho}\Bigr)
{{d\ell}\over {dt}},
\end{equation}
\noindent 
where $\rho$ is the density of the object.

For small particles such as micron-sized and submicron-sized
objects, one can replace $|{\bf V}-{\bf U}|$ with
${{\bf V}_{\rm {rel}}}$, the relative
velocity of the object with respect to the gas.
The radial, vertical and tangential
components of this velocity are, respectively, given by 
${\dot r},\, {\dot z}$, and $r({\dot \varphi}\,-\,{\omega_g})$,
where the motions of gas molecules along the $z$-axis 
have been neglected. Because the gas is at hydrostatic
equilibrium, its angular velocity,
$\omega_g$, is slightly different from its Keplerian
value and is given by
\begin{equation}
{\omega_g^2} ={{GM}\over {({r^2}+{z^2})^{3/2}}} \,+\,
{1\over {r{\rho_g}(r,z)}}\,
{{\partial{{\cal P}_g}(r,z)}\over {\partial r}}.
\end{equation}
\vskip 5pt
\noindent
In this equation, 
${{\cal P}_g}(r,z)={K_B}T{\rho_g}(r,z)/{m_H}$ 
is the pressure of the gas.

An object in the model nebula considered here is subject
to the gravitational attraction of the central star and
the drag force of the gas.
The equation of motion of an object in this nebula
is, therefore, given by
\begin{equation}
m{\ddot{\bf R}}\,=\,
-\,{{GMm}\over {({r^2}+{z^2})^{3/2}}}\,{\bf R}\,-\,
\pi\, {\rho_{\rm{dust}}}(r,z)\,
{a^2}\,{{d\ell}\over {dt}}\, {\bf V}_{\rm {rel}}\,-\,
{{\bf F}_{\rm {drag}}}\,,
\end{equation}
where ${\bf R}(r,z)$ is the position vector of the particle
and
\begin{equation}
{\bf F}_{\rm {drag}}= {{2{a^2}}\over{3(a+\lambda)}}
{\biggl({{2\pi{K_B}T}\over{{m_H}}}\biggr)^{1/2}}
\Bigl[\lambda{\rho_g}(r,z)+{{3{m_H}}\over {2\sigma}}\Bigr]
{{\bf V}_{\rm {rel}}}
\end{equation}
\noindent
represents the drag force of the gas. In equation (7),
$\sigma$ is the collisional cross section between two hydrogen
molecules and $\lambda$ is their mean free path.

\section{Numerical Results}

The equation of motion of a particle and the
growth equation (4) were integrated, numerically,
for objects with initial radii ranging from 
1 to 100 microns. The mass of the central star
was chosen to be one solar mass,
$\beta =1\,,{r_m}=1$ AU, and
${\rho_0}={10^{-9}}$g cm$^{-3}$. The collisional
cross section of hydrogen molecules, $\sigma$, was
taken to be $2\times {10^{-15}}\,{\rm {cm}}^{-2}$,
and their mean free paths 
$\lambda\,{\rm {(cm)}}=
4\times {10^{-9}}/{\rho_g}(r,z)\,({\rm {g cm}}^{-3})$.
At the beginning of integrations, an object was placed
at $(r,r/10)$, with a Keplerian circular radial velocity and
with no motion along the $z$-axis.
Figure 1 shows the growth of two one micron-sized objects, one
migrating radially inward from (2,0.2) AU, and one migrating 
radially outward from (0.25, 0.025) AU to (1,0) AU, 
the location of the maximum gas
density on the midplane. For the comparison, the radial
and vertical motions of these objects without mass-growth
have also been plotted. As shown here, by sweeping up the
smaller particles of the background, these objects grow to 
a few centimeters in size and approach the midplane
in a time much shorter than the time of similar
migration without the mass-growth.

Equations (4) and (6) were also integrated for different 
values of the object's density and gas temperature.
As expected, the rate of growth of the object increased
with an increase in the temperature of the gas (Fig. 3). This
can be attributed to larger value of the pressure gradients
at higher temperatures \cite{hb03a}. Increasing the density of the object
while keeping the temperature of the gas constant had, however,
an opposite effect. Objects with higher densities tend to 
grow in size over longer periods of time. The inverse proportionality
of the rate of the growth of the object to its density
can also be seen from equation (4).

\begin{figure}
  {\includegraphics[height=.3\textheight]{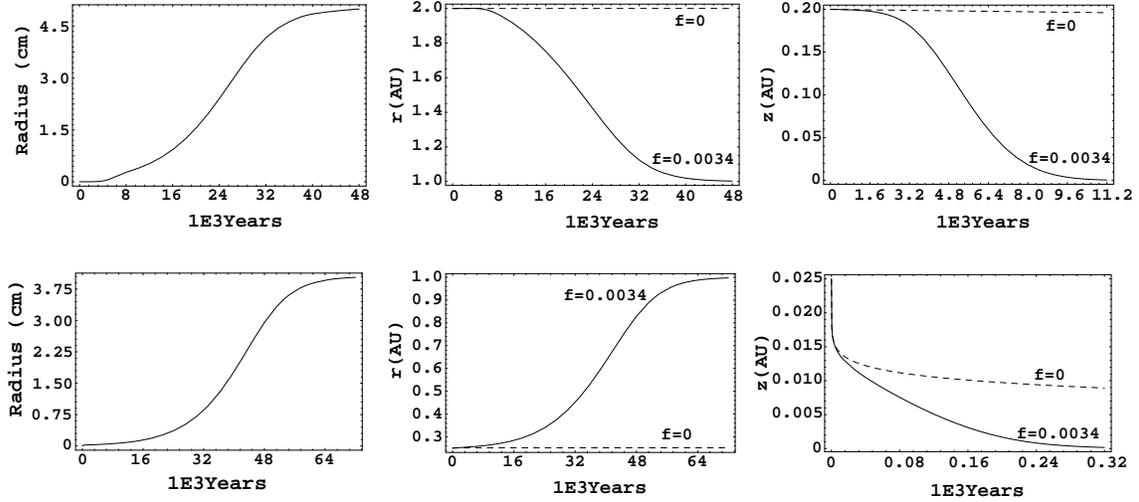}}
  \caption{Graphs of the radius (left), radial migration (middle),
           and vertical descent (right) of a 1 micron object with
           a density of 2 g/cc. The gas temperature is 300 K.
           The dashed line indicates migration without mass-growth.
           The middle graphs show an inward radial migration from 
           2 AU (top), and an outward radial migration from 0.25
           AU (bottom) to the location of the maximum gas density
           on the midplane, (1,0) AU. The initial value of $z$ 
           was taken to be $r/10$.}
\end{figure}

\begin{figure}
  {\includegraphics[height=.3\textheight]{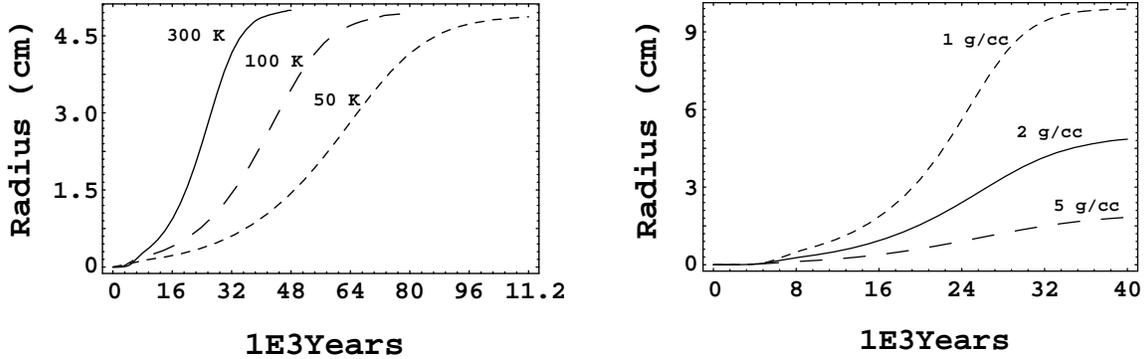}}
  \caption{The growth of a 1 micron object with a density of
           2 g/cc for different values of the gas temperature
           is shown on the left. The right graph depicts the
           growth of similar particle with different densities
           in an isothermal gas with a temperature of 300 K.
           The particle was initially at (2,0.2) AU.}
\end{figure}


\begin{theacknowledgments}
This work is partially supported by the NASA Origins of 
the Solar System Program under
grant NAG5-11569, and also the NASA Astrobiology Institute
under Cooperative Agreement NCC2-1056.
\end{theacknowledgments}

\bibliographystyle{aipproc}

\end{document}